\documentclass[aps,prl,twocolumn,groupedaddress]{revtex4}
\usepackage{amsmath,amssymb,graphicx,dcolumn,times,bm}
 \usepackage[dvips]{color} 
\def\br{{\mathbf{r}}}
\def\exS{{S^\mathrm{ex}}}
\begin{document}


\title{Tuning density profiles and mobility of
  inhomogeneous fluids}

\author{Gaurav Goel}
\affiliation{Department of Chemical Engineering, The University of
  Texas at Austin, Austin, Texas 78712} 
\email{goel@che.utexas.edu, krekel@che.utexas.edu}

\author{William P. Krekelberg} 
\affiliation{Department of Chemical Engineering, The University of
  Texas at Austin, Austin, Texas 78712}
\email{krekel@che.utexas.edu}

\author{Jeffrey R. Errington} 
\affiliation{Department of Chemical and Biological Engineering,
  University at Buffalo, The State University of New York, Buffalo,
  New York 14260} 
\email{jerring@buffalo.edu}

\author{Thomas M. Truskett} 
\affiliation{Department of Chemical Engineering and Institute for
  Theoretical Chemistry, The University of Texas at Austin, Austin,
  Texas 78712}
\email{truskett@che.utexas.edu}



\date{\today}

\begin{abstract}
  Density profiles are the most common measure of inhomogeneous
  structure in confined fluids, but their connection to 
transport coefficients is poorly understood.  We explore via
  simulation how tuning particle-wall interactions to flatten or enhance
  the particle layering of a model confined fluid impacts its
  self-diffusivity, viscosity, and entropy.  Interestingly,
  interactions that eliminate particle layering significantly reduce
  confined fluid mobility, whereas those that enhance layering can 
  have the opposite effect. Excess entropy helps to understand and
  predict these trends.
\end{abstract}


\maketitle


Fluids confined to narrow spaces adopt an inhomogeneous
distribution of density due to the interactions between the fluid
particles and the boundaries.  This density profile provides a basic 
means for characterizing confined fluid
structure, and in particular how it differs from that of a homogeneous bulk
system.  For simple confined fluids, the density profile can be
quantitatively predicted using classical density functional theory.
However, differences between confined and bulk fluids 
are not limited to
static structure. The former also flow, diffuse, and conduct
heat at different rates than the latter. Unfortunately, a theory
that can reliably predict transport coefficients has yet to emerge.  In
fact, even an intuitive understanding of how the density
profile of a confined fluid connects to its dynamics is lacking.

For bulk fluids, semi-empirical structure-property relations have
helped to correlate and predict transport coefficients (see, e.g., \cite{Rosenfeld1977Relationbetweentransport,Rosenfeld1999quasi-universalscalinglaw,Dzugutov1996universalscalinglaw}). Specifically,
changes in thermodynamic state variables that increase short-range
structural order of fluids are also known to 
decrease their mobility in a
simple, quantifiable way.  This is true even for systems that
exhibit anomalous dynamical behavior, such as cold liquid water (where
viscosity decreases upon compression)~\cite{Errington2006Excess-entropy-basedanomalieswaterlike,Mittal2006QuantitativeLinkbetween,Sharma2006Entropydiffusivityand} or concentrated colloidal
suspensions (where interparticle attractions increase mobility)~\cite{Krekelberg2007Howshort-rangeattractions,Mittal2006QuantitativeLinkbetween}.
Na{\"{i}}ve extrapolation of this idea might lead one to suspect that inhomogeneous 
fluids with highly structured (e.g., layered) density profiles would tend to be more viscous
and less diffusive than more spatially uniform fluids.  Is that
indeed the case?
Here, we
explore this issue quantitatively. Specifically, we use 
molecular simulation to 
investigate the relationship between the transport coefficients of an 
inhomogeneous fluid and its density profile, the latter of which can be 
modified in a precise way through the interactions of the fluid particles 
with the confining boundaries.          

A key empirical observation motivating this study is the existence of an
isothermal correlation between the self-diffusion
coefficient of simple inhomogeneous fluids and excess entropy
(relative to ideal gas), which
is approximately obeyed across a wide range of confining
environments~\cite{Mittal2006ThermodynamicsPredictsHow,Mittal2007Doesconfininghard-sphere,Mittal2007RelationshipsbetweenSelf-Diffusivity}.
Since the magnitude of the excess entropy is itself a
measure of structural order~\cite{Truskett2000Towardsquantificationof}
, the
aforementioned correlation is effectively a structure-property relationship.  
But how does excess entropy connect to the density profile?
Do fluids with more structured density profiles generally have 
lower or higher values of excess entropy when compared to spatially
uniform fluids?  
Moreover, can excess entropy be tuned via the fluid-boundary interactions to 
modify the transport coefficients in a controlled and predictable way?
If so, this idea might be used to great effect in the engineering of
micro- and nano-fluidic systems.

To address these open questions, we turn to the well-characterized 
Weeks-Chandler-Andersen (WCA) model~\cite{chandler:83}, which is 
known to capture the entropic
packing effects that control many properties of 
dense, atomistic and colloidal fluids. 
The WCA pair potential is defined as
$\phi(r)=4\epsilon([\sigma/r]^{12}-[\sigma/r]^{6})+\epsilon$ for
$r<2^{1/6}\sigma$ and $\phi(r)=0$ for $r>2^{1/6}\sigma$, where 
$r$ is the interparticle separation.   We consider this fluid
confined to 
a thin film geometry between 
two parallel, planar boundaries
placed a distance $H$ apart.  Particles located a 
distance $z$ from one boundary ($0<z<H$) interact with 
an external field
$\phi_{\mathrm{ext}}(z)=\phi_{\mathrm{fw}}(z)+\phi_{\mathrm{fw}}(H-z)$. The
    single-wall potential is given by
    $\phi_{\mathrm{fw}}(z)=(2/15)[\sigma/z]^{9}-[\sigma/z]^{3}+\sqrt{10}\epsilon/3$ +
    $\phi_0(z)$ for $z<(2/5)^{1/6}\sigma$ and
    $\phi_{\mathrm{fw}}(z)=\phi_0(z)$ for $z>(2/5)^{1/6}\sigma$.  This
    represents a WCA 9-3 repulsive boundary plus an additional term
$\phi_0(z)$, which can be used to tune 
the density profile. 
From here forward, we simplify notation by reporting quantities implicitly
nondimensionalized by appropriate
combinations of the characteristic length scale $\sigma$ and 
energy scale $\epsilon$ (or equivalently $k_{\mathrm{B}}T$, since 
we set $\epsilon=k_{\mathrm{B}}T$ for all calculations).  
In the above, $k_{\mathrm{B}}$ is the Boltzmann constant and $T$ is
temperature.  

Our aim is to investigate, for given values of film thickness 
$H$ and average fluid density 
$\rho_{\mathrm{avg}}=H^{-1}\int_0^{H}\rho(z)dz$, how the details of the 
density profile $\rho(z)$ impact relaxation processes.  
This can be done systematically if a suitable set of target density
profiles can be chosen for study.
For fixed chemical potential $\mu$, $H$, and $T$, there is
a one-to-one mapping between $\phi_0(z)$ and $\rho(z)$~\cite{Weeks2003ExternalFieldsDensity}.    
In fact, as explained below, the specific $\phi_0(z)$ that will produce a
given target density profile can be determined precisely using
Monte Carlo (MC) simulation methods.  
Once determined, $\phi_0(z)$ can then be imposed in an 
equilibrium molecular dynamics simulation to calculate the transport
coefficients of the fluid.

The baseline density profile that we consider 
is the ``natural'' one for the WCA fluid confined between WCA 9-3
walls, i.e., the profile adopted by the equilibrium fluid when $\phi_0(z)=0$.  It is characterized 
by a moderately inhomogeneous structure 
of fluid particles layered parallel to the confining boundaries.

\begin{figure}[h]
   \includegraphics{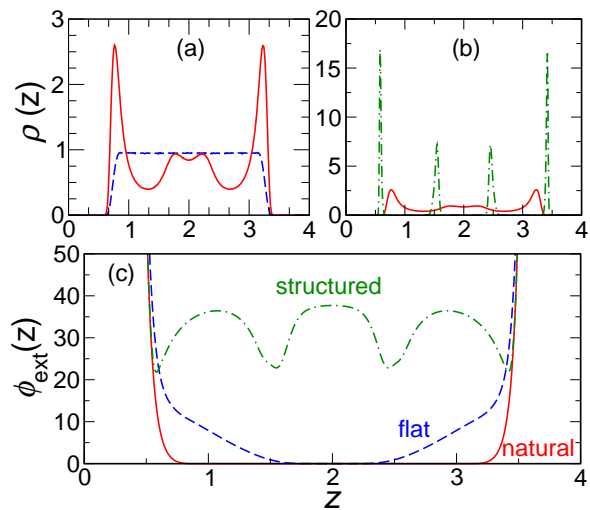}%
   \caption{\label{fig1_rhoz_potz} (a) Natural and flat density
     profiles $\rho (z)$, and (b) natural and structured density
     profiles for a confined WCA fluid with average density
     $\rho_\mathrm{avg}=0.6$ and $H=4$, as discussed in the text. (c) The
     associated particle-boundary
     interactions $\phi_{\mathrm{ext}}(z)$.}
 \end{figure}
A second type of profile that we consider is a ``flat'' density
distribution, where the layered structure of the confined 
WCA fluid is effectively eliminated by judicious choice
of $\phi_0(z)$.  We define this flat profile to be equivalent to that 
which an equilibrium fluid of non-interacting particles with average density 
$\rho_{\mathrm{avg}}$ would adopt in 
the presence of the boundary potential $\phi_{\mathrm{ext}}(z)$
with $\phi_0(z)=0$. 
Fig.~\ref{fig1_rhoz_potz}(a) compares the shapes of typical 
flat and natural density
profiles, and Fig.~\ref{fig1_rhoz_potz}(c) shows the external potentials 
$\phi_{\mathrm{ext}}(z)$ that produce them for the WCA fluid.  
Interestingly, 
due to the non-local coupling of $\phi_{\mathrm{ext}}(z)$ to 
$\rho(z)$, the layering structure of WCA
particles in the natural
profile can be generally eliminated by the addition of a non-oscillatory, 
repulsive contribution to the external field.
As we demonstrate below, flattening the density profile in this way 
has the effect of reducing both the entropy and the excess 
entropy of the confined fluid.

One can imagine that $\phi_0(z)$ could alternatively be
chosen such that the confined 
fluid takes on a higher excess entropy (and, thus, perhaps
a higher mobility) than when $\phi_0=0$ (i.e., 
the natural density profile). 
To demonstrate this, we investigate a third class of density 
distributions which we refer
to as ``structured''.  The procedure for generating structured
density profiles is described in detail in the Appendix.  In short, they 
correspond to confined fluid states predicted by 
an approximate density functional theory to have 
particularly high values of excess entropy.  
As is evident in Fig.~\ref{fig1_rhoz_potz}(b), structured profiles do 
have more pronounced layering than their natural counterparts, an {\em
  a posteriori} justification for their name.  
Moreover, the MC simulations described below verify that 
they indeed correspond to states with 
higher excess entropy 
(but lower entropy) than those with the natural profile.

Here, excess entropy is defined in the usual way as the
difference between the fluid's entropy and that of an ideal gas with
the same density profile.  
To obtain excess entropy data for the aforementioned systems, 
we employed a suite of MC
techniques. 
First, we 
used grand canonical transition matrix MC
simulations~\cite{Errington2003} with volume $V=1000$ to
determine excess entropy as a function of $N$ 
(or equivalently $\rho_{\mathrm{avg}}$) and $H$ for the natural
profile 
case.  
A detailed explanation of this approach can be found in~\cite{Mittal2006ThermodynamicsPredictsHow,Errington2006Excess-entropy-basedanomalieswaterlike}.  
For select values of $N$ and $H$, we then determined the $\phi_0(z)$
that generated a target (flat or structured) density profile.  We 
subsequently used an expanded ensemble MC~\cite{Lyubartsev1992Newapproachto} procedure to determine
the excess entropy of a fluid subjected to this potential. 

To compute $\phi_0(z)$, we used an efficient nonequilibrium MC potential 
refinement technique recently 
introduced by Wilding~\cite{Wilding2003nonequilibriumMonteCarlo}.  
The potential was initialized with $\phi_0(z)=0$ and subsequently
tuned during an eight-stage canonical MC simulation.
At regular 
intervals during the $i$th stage, $\phi_0(z)$ was
incremented 
by the relative difference between the instantaneous and target
density profiles scaled by a modification factor $y_i=0.001/2^i$.  Each stage
terminated when $\zeta$, the maximum relative difference between the 
target and aggregated stage density profiles, dropped below a
tolerance of $\zeta^*=0.01$.

Thermodynamic properties of the confined WCA fluid subjected to a 
nonzero target $\phi_0(z)$ were determined through canonical expanded
ensemble simulations~\cite{Lyubartsev1992Newapproachto}.  A system with a flat or structured profile 
was related to that with a natural profile through a series of
subensembles in which the target potential $\phi_0(z)$ was scaled by a factor 
$\lambda$ that spanned from 0.0 to 1.0 in increments of 0.001.  
A transition matrix MC technique similar to that described in~\cite{Errington2007Calculationofsurface} was
used to determine the relative Helmholtz free energy $F(\lambda)$ of
each subensemble.  The average energy $E(\lambda)$ was also accumulated 
during a simulation.  The total entropy difference 
$\Delta S(\lambda)= S(\lambda)-S(0)$  between a subensemble
characterized by $\lambda$ and a fluid with a natural profile was
evaluated as $\Delta
S(\lambda)=T^{-1}\left([E(\lambda)-E(0)]-[F(\lambda)-F(0)]\right)$.  
The corresponding difference in the excess entropy  $\Delta
S^{\mathrm{ex}}(\lambda)$ can be written as $\Delta
S^{\mathrm{ex}}(\lambda)=\Delta S(\lambda) - \Delta
S^{\mathrm{id}}(\lambda)$.  The term $\Delta
S^{\mathrm{id}}(\lambda)=-VH^{-1}\int_0^H
\left[\rho(z,\lambda)\ln\rho(z,\lambda)-\rho(z,0)\ln\rho(z,0)\right]
dz$ is simply the change in entropy of an ideal gas upon changing 
its density profile from $\rho(z,0)$ to $\rho(z,\lambda)$. 

We computed the transport coefficients of the thin films via 
molecular dynamics simulations in the microcanonical ensemble using $N=4000$
particles and integrating the equations of motion with the
velocity-Verlet algorithm~\cite{allen89ComputerSimulationOf}.  A time step of 0.0025 was used for
simulating the
natural and flat profile systems, while a shorter time step of 0.0002 was
employed for the structured profile fluids.   
Periodic boundary conditions were employed in the $x$ and
$y$ ``free'' directions. We
extracted values of self-diffusivity (parallel to the walls) $D$ by fitting the 
long-time ($t>>1$) behavior of the mean-squared displacement to 
the Einstein relation for diffusion
$\big<\Delta \mathbf{r}^2\big>=4Dt$, where
$\big<\Delta \mathbf{r}^2\big>$ corresponds to the
mean-squared displacement in the $x$ and $y$ directions. 
We also calculated values of zero-shear viscosity $\eta$ using its 
corresponding Einstein relation.

In order to systematically probe the effects of 
flattening or enhancing the layering of the density profile, we
first focus on the behavior of the confined fluid for
$\rho_{\mathrm{avg}}=0.6$ and $H=4$.  Let
$\phi_{0,{\mathrm{f}}}(z)$ and $\phi_{0,{\mathrm{s}}}(z)$ 
represent the contributions
to the external potential that, under these conditions, produce the 
flat and structured
profiles, respectively.  Starting from the 
natural profile, we incrementally flatten the density distribution 
by choosing
$\phi_0(z) = \lambda_{\mathrm{f}} \phi_{0,{\mathrm{f}}}(z)$
with progressively larger values of $\lambda_{\mathrm{f}}$ in the range 
$0 \le \lambda_{\mathrm{f}} \le 1$.  Similarly, we systematically 
enhance the layering of the natural profile by setting $\phi_0(z) =
\lambda_{\mathrm{s}} \phi_{0,{\mathrm{s}}}(z)$ with
progressively larger values of $\lambda_{\mathrm{s}}$ in the range 
$0 \le \lambda_{\mathrm{s}} \le 1$.

\begin{figure}[h]
   \includegraphics{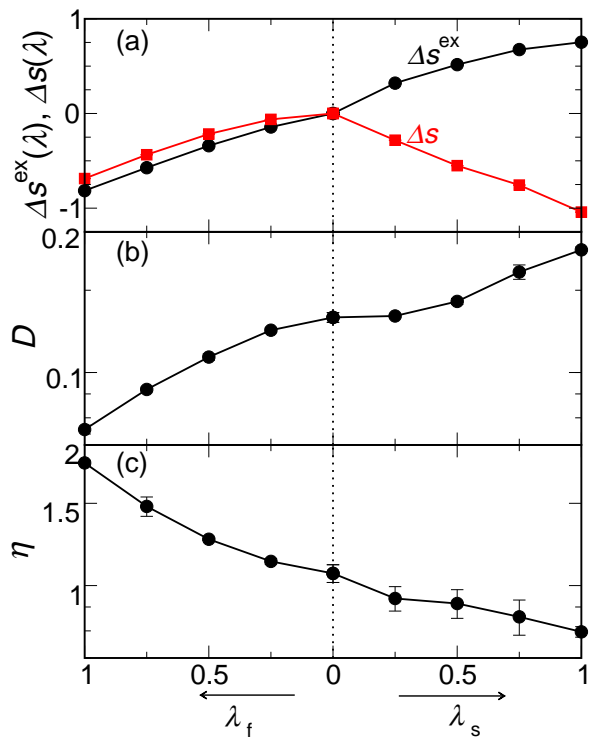}%
   \caption{\label{fig2_lambdavary} Effect of boundary interaction
     (shape of density profile)
     on (a) excess entropy per particle 
$\Delta s^\mathrm{ex} (\lambda) = s^\mathrm{ex} (\lambda)-
s^\mathrm{ex} (0)$ and entropy per particle $\Delta s(\lambda) = s(\lambda)-
s(0)$, (b)
     self-diffusivity $D$, and (c) viscosity $\eta$ for the confined
     WCA fluid  with
     $\rho_{\mathrm{avg}}=0.6$ and $H=4$.  The centerline
     corresponds to the fluid with the natural density profile of 
Fig. 1(a) and (b).   From center to left, the density profile is
systematically flattened:  $\phi_{0}(z)=\lambda_{\mathrm{f}}
     \phi_{0,{\mathrm{f}}}(z)$, where  $\lambda_{\mathrm{f}}=1$
     yields the flat profile shown in Fig.~1(a). From center to right, the
     density profile is structured:  $\phi_{0}(z)=\lambda_{\mathrm{s}}
     \phi_{0,{\mathrm{s}}}(z)$, where  $\lambda_{\mathrm{s}}=1$
     produces the structured profile shown in Fig.~1(b).  
Symbols are simulation data,
     and curves are guide to the eye.}
 \end{figure}
In Fig.~\ref{fig2_lambdavary}, we show how these specific ways of
modifying the density distribution in turn affect the entropy per particle $s$, the excess entropy per
particle $s^{\mathrm{ex}}$, the self-diffusivity $D$, and the viscosity
$\eta$.  As expected, $s$ appears highest for the natural profile.  This is 
because the system is virtually athermal when $\phi_0=0$
due to the steep boundary and 
interparticle repulsions.  As
a result, its equilibrium (minimum free energy) structure also 
approximately maximizes
$s$ relative to other fluid states [i.e., other  $\phi_0(z)$ and
corresponding density profiles] with the same
$\rho_{\mathrm{avg}}$ and $H$.  The natural profile state does not,
however, maximize $s^{\mathrm{ex}}$.  
Rather, $s^{\mathrm{ex}}$ is found to monotonically increase with increased 
``structuring'' of the density profile. 
How can we understand these trends?  

The key is to recall that 
$\Delta s^{\mathrm{id}}(\lambda)=\Delta s(\lambda)-\Delta
s^{\mathrm{ex}}(\lambda)$ quantifies how changing $\lambda$ and, in
turn, the density profile modifies the ideal gas entropy, while 
$\Delta s^{\mathrm{ex}}(\lambda)$ measures the corresponding entropic change
associated with the interacting fluid's interparticle correlations.  
With this is mind,
it is clear that flattening the natural profile will
inevitably result in an increase in ideal gas entropy [i.e., $\Delta s^{\mathrm{id}}(\lambda_{\mathrm{f}})>0$].
Moreover, since $s$ generally decreases upon flattening of the natural 
profile, it follows that $s^{\mathrm{ex}}$ will also
 generally decrease, reflecting an overall strengthening of interparticle
 correlations.  In other words, paradoxically, 
the fluid with the flat density profile
 exhibits the highest degree of structural order.   

On the other hand, it is also self-evident that enhancing the
 layering of the natural density profile will decrease the ideal gas entropy
 [i.e., $\Delta s^{\mathrm{id}}(\lambda_{\mathrm{s}}) < 0$].  
Since $s$ also decreases in this process, the aforementioned 
entropic penalty can at most be 
partially compensated by an increase in
  $s^{\mathrm{ex}}$ due to diminished interparticle correlations.  
In fact, Fig.~\ref{fig2_lambdavary} provides a specific example for how
``structuring'' the density profile can significantly 
decrease the overall structural order (i.e., increase $s^{\mathrm{ex}}$) of
the confined fluid.  One should keep in mind that 
the specific structured profiles examined here 
were selected precisely because density functional theory predicted
that they would have high values of $s^{\mathrm{ex}}$.  Thus, we emphasize that
while flattening the natural density profile can generally be expected
to decrease $s^{\mathrm{ex}}$, alternative means for embellishing its 
structure at constant $\rho_{\mathrm{avg}}$ 
may either increase or decrease $s^{\mathrm{ex}}$. The case
studied here provides proof of concept for the former. We will 
extensively discuss the latter case in a future publication.  

A second important point of Fig.~\ref{fig2_lambdavary} is that the 
mobility of the confined fluid, as measured by
both $\eta^{-1}$ and $D$, closely tracks the
behavior of $s^{\mathrm{ex}}$ (but not $s$) for the flattening and structuring
processes described above.  A physical explanation is that
fluid transport parallel to the walls is dominated by interparticle
collisions, and hence structural
correlations~\cite{Rosenfeld1977Relationbetweentransport,Rosenfeld1999quasi-universalscalinglaw,Dzugutov1996universalscalinglaw}.
The density profile appears to
impact these transport processes mostly due to the fact that it
modifies the 
interparticle correlations,
the effect of which is conveniently isolated by computing the 
excess rather than 
the total entropy of the fluid.

The above observation that fluids with more uniform density profiles
can actually have lower $s^{\mathrm{ex}}$ and slower dynamics than those
with strongly layered density profiles (and the same $\rho_{\mathrm{avg}}$) 
appears even more general when viewed in the context of other recent
simulation data.  In particular, it is now known that the hard-sphere
fluid confined between hard walls shows lower $s^{\mathrm{ex}}$~\cite{Mittal2007Doesconfininghard-sphere} and
slower single-particle dynamics both parallel~\cite{Mittal2007Doesconfininghard-sphere} and
normal~\cite{mittal2008} to the confining walls
when such walls are separated by distances that inherently frustrate 
the ability of the fluid to structure into an integer number of layers in 
its density profile.  These model predictions of the relationship 
between dynamics and
density profiles can now also be readily tested in 
experiments, e.g., by using confocal microsopy to investigate confined 
``hard-sphere'' colloidal suspensions~\cite{Weeks2000Three-DimensionalDirectImaging}.  

\begin{figure}[h]
   \includegraphics{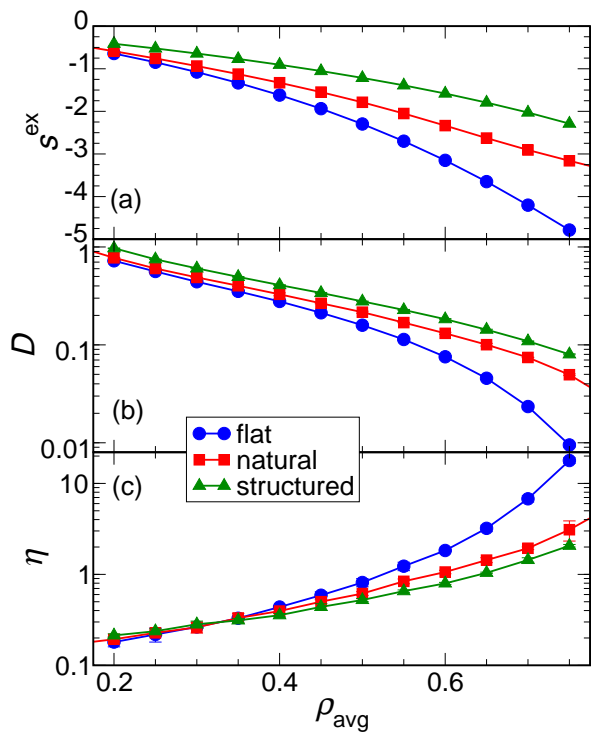}%
   \caption{\label{fig3_DetaexS} (a) Excess entropy per
     particle $s^\mathrm{ex}$, (b) self-diffusivity $D$, and (c) viscosity
     $\eta$, of the confined WCA fluid versus average density
     $\rho_{\mathrm{avg}}$ at $H=4$.  Symbols are simulation data,
     and curves are guide to the eye.}
 \end{figure}
Finally, to document the generality of the physics discussed above, 
we show in Fig.~\ref{fig3_DetaexS} the behaviors
of  $s^{\mathrm{ex}}$,  $\eta$, and $D$ over a broad range of
average film densities.  Perhaps most striking is the observation
that, at high density,
fluids with structured and flat density profiles with the 
same $\rho_{\mathrm{avg}}$
can differ in both $D$ and $\eta$ by an order of
magnitude. Although the current study focuses on equilibrium fluid conditions,
the trends evident in Fig.~\ref{fig3_DetaexS} suggest that one might
even be able to effectively supercool monatomic 
confined fluids by isothermally modifying the external potential in a
way that systematically flattens the density profile.  Finally, one
can imagine extending these ideas to confined mixtures by using a 
species-dependent boundary potential to tune the 
partial molar entropies and dynamics of individual components.
We are
currently exploring both of these possibilities.

%
%
\begin{acknowledgments}
T.M.T. and J.R.E. acknowledge support of the National
Science Foundation (NSF) under Grant Nos. CTS-0448721 and CTS-028772,
respectively. T.M.T. acknowledges support of the David and
Lucile Packard Foundation and the Alfred P. Sloan
Foundation. W.P.K. and G.G. acknowledges support from a NSF Graduate Research
Fellowship and a UT ChE department fellowship, respectively. The Texas Advanced
Computing Center (TACC) and University at Buffalo Center for
Computational Research provided computational resources for this study.
\end{acknowledgments}

\renewcommand{\theequation}{A-\arabic{equation}}
\setcounter{equation}{0}  
\begin{center}
    {\bf APPENDIX}
  \end{center}
\section*{Protocol for generating high excess entropy density profiles}
Here, we explain how the
so-called ``structured'' target density profiles in our study were obtained.
As discussed in the main text, the idea underlying the structured
profile is that the corresponding confined equilibrium fluid [with
non-zero $\phi_0(z)$] should have a high excess
entropy relative to the value it takes on when it adopts the natural 
profile [$\phi_0(z)=0$] at the same $H$ and
$\rho_{\mathrm{avg}}$.  Clearly, the name ``structured'' does not follow in any
logical sense from the above, but it was rather chosen {\em a posteriori}
to qualitatively 
reflect the shapes of the profiles that we find have this property.

So how does one search out density profiles associated 
with high excess
entropy of the fluid?  The approach we adopt here is to use classical density functional
theory (DFT) for inhomogeneous fluids, which provides a formal, although
necessarily approximate, connection between excess entropy and the
density profile.  Specifically, we make use of a recent
modification~\cite{Yu2002Structuresofhard-sphere} of Rosenfeld's accurate
fundamental measure theory~\cite{Rosenfeld1989Free-energymodelinhomogeneous} for confined
hard-sphere fluids.  Our simulated confined WCA fluid can be
accurately mapped onto a confined hard-sphere (HS) fluid with the same
effective packing fraction $\pi\rho_{\mathrm{avg}}
\sigma_{\mathrm{HS}}^3/6$ by assigning the
WCA particles an effective, HS interparticle diameter
$\sigma_{\mathrm{HS}}$.  This diameter satisfies the following
``Boltzman factor
criterion''~\cite{Ben-Amotz2004ReformulationofWeeks-Chandler-Andersen},
$\phi(r=\sigma_{\mathrm{HS}})=1$, where $\phi(r)$ is the WCA
interparticle pair potential given in the letter.   

In the context of DFT, the excess entropy of a confined HS fluid can
be expressed as, $\exS[\rho(\br)]=- {\cal{F}^{\mathrm{ex}}[\rho(\br)}]=-\int d\br \Phi (\br)$,
where $\Phi (\br)$ is the excess intrinsic Helmholtz
free energy density.  One can further write down an Euler-Lagrange  
equation for maximizing excess entropy, 
\begin{equation}
  \label{eq:EL_general}
\frac{\delta \exS[\rho(\br)]}{\delta
  \rho (\br)}=-\frac{\delta {\cal{F}}^{\mathrm{ex}}[\rho(\br)]}{\delta
  \rho (\br)}= c^{(1)} (\br)=0.
\end{equation}
In the above, the second equality follows from the definition of the
one-body direct correlation function $c^{(1)} (\br)$.  But how can this relationship be understood?  

From the potential distribution theorem~\cite{Widom1963SomeTopicsin,Hendersen1983StatisticalMechanicsOf}, 
we have $c^{(1)} (\br)=\ln p_0(\br)$, where $p_0(\br)$ is the probability of
inserting an additional hard-sphere at position $\br$ without
overlapping a particle of the fluid.  From this, it is clear that 
$c^{(1)} (\br)=0$ cannot generally be satifisfied for
finite $\rho_{\mathrm{avg}}$.  However, for our present purposes, we are not
interested in finding states that strictly maximize excess entropy.  
Rather, we want our target ``structured'' profiles to exhibit high values of
excess entropy relative to that of the natural density profile.
Fortunately, for that purpose, Eq.~\ref{eq:EL_general} can still be
productively used if recast in a different form. 

Specifically, the 
equation ${\delta \exS [\rho(\br)]}/{\delta \rho (\br)}=0$
with the constraint of constant $\rho_{\mathrm{avg}}$ can be expressed as
\begin{equation}
  \label{eq:eulerlagrange_exs}
  \dfrac{\rho (z)}{\rho_\mathrm{avg}}=\dfrac{\rho (z) \exp [c^{(1)}
    (z)]}{H^{-1} \int_0^H dz \rho (z) \exp [c^{(1)}(z)]}
\end{equation}
for the slit-pore geometry.  Inputting a trial $\rho (z)$ and
the associated  $c^{(1)}(z)$ predicted from fundamental measure theory
into the right hand-side of Eq.~\ref{eq:eulerlagrange_exs} produces
a new $\rho (z)$ on the left-hand side.  We observed that 
thus performing Picard style iterations on
Eq.~\ref{eq:eulerlagrange_exs} eventually 
yields density profiles that correspond
to fluid states with higher excess entropy than that of 
the natural profile. In particular, we carried out these iterations
with a mixing parameter~\cite{mixparnote} $\kappa=0.005$ 
until the relative change in $\exS$
(=$\kappa^{-1} (\exS_{n+1}-\exS_{n})$) was less than
$0.05$.  Weighted densities and integrals were calculated using the
trapezoidal rule on a mesh of $0.005\sigma_\mathrm{HS}$.  For the highest
density case, $\rho_\mathrm{avg}=0.75$, we used the
corresponding natural density
profile as our initial guess.  For all lower densities, we used the
structured density profile at $\rho_\mathrm{avg}=0.75$ as the initial
guess.    

All target density profiles and corresponding external potentials used
in this study are available upon request.


\begin{thebibliography}{21}
\bibitem[{mix()}]{mixparnote}
\bibinfo{note}{To avoid divergence, we mix the input density
  $\rho_\mathrm{in}^{n}(z)$ (right-hand side of
  Eq.~\ref{eq:eulerlagrange_exs}) and the output density
  $\rho_\mathrm{out}^{n}(z)$ (left-hand side of
  Eq.~\ref{eq:eulerlagrange_exs}) 
  for iteration $n$ to obtain the input density for iteration $n+1$, i.e.,
  $\rho_\mathrm{in}^{n+1} (z) = \rho_\mathrm{out}^{n} (z)\kappa +
  \rho_\mathrm{in}^{n} (z) (1-\kappa)$.}
\end{thebibliography}
\end{document}